\documentclass[12pt,a4paper]{article}
\usepackage{amsmath}
\usepackage[cp1251]{inputenc}
\usepackage[english]{babel}
\usepackage{amsfonts}
\usepackage{amsfonts,amssymb}
\usepackage{amssymb}
\usepackage{latexsym}
\usepackage{euscript}
\usepackage{enumerate}
\usepackage{graphics}
\usepackage[dvips]{graphicx}
\topmargin = - 1 cm
\begin{document}
\large

\begin{center}
{\bf Cluster decomposition of percolation probability\\
on the hexagonal lattice}

Antonova E.S., Virchenko Yu.P.\\
Belgorod State University, Russia
\end{center}

{\small The upper estimate of the percolation threshold of the
Bernoulli random field on the hexagonal lattice is found. It is
done on the basis of the cluster decomposition. Each term of the
decomposition is estimated using the number estimate of cycles
on the hexagonal lattice which represent external borders of possible
finite clusters containing the fixed lattice vertex.}
\bigskip

{\bf 1. Introduction.} Generally, the percolation theory studies
random subsets of the infinite set where the filtering relative to
the inclusion and the supplementary connectedness relation has
been defined \cite{VirE}. In particular, such a theory may be
arisen in noncompact topological space where filtering is defined
by sequences of compact spaces when their union coincides with the
total space \cite{VirAlg}. The problem of existence of the random
realization with connected noncompact components represents the
main interest. In the case when the probability of such an event
is positive, they say that the percolation is present in the
random set. The calculation of the percolation probability
represents the difficult  mathematical problem even in most simple
mathematical structures pointed out. Therefore, they resort
usually to computer ex\-pe\-ri\-ments for the problem solving when
the percolation theory is applied and there are some considerable
achievements in this direction (see, for example, \cite{Stau}). In
this work, we study the pointed out problem from the mathematical
point of view. Usually, due to the extreme complexity of the
problem, such investigations are connected with the study of
elementary noncompact spaces  such as integer lattices $ {\Bbb Z}
^d $, $d = 2, 3$ with the definite connectedness relation. The
definition of this relation transforms integer lattices into those
mathematical objects which are called periodic graphs \cite {1}.
Percolation on periodic graphs is the subject of {\it the discrete
percolation theory}. However, even for periodic graphs, the main
problem of percolation theory are resisted to mathematical
processing only for random sets generated by the Bernoulli field
$\{{\tilde c}({\bf x}); {\bf x} \in {\Bbb Z}^d\}$ when the
probability distribution defined by the unique parameter $c = {\rm
Pr} \{{\tilde c} ({\bf x}) = 1 \} $. In this work, it is found the
upper estimate of the so-called {\it percolation threshold} $c_
* $ in the case of the two-dimensional uniform periodic
graph which is called the hexagonal lattice. The percolation
probability $Q (c)$ is differed from zero at $c> c_*$. Our
estimate is done by the well-known approach  which we name
{\it the cluster decomposition} (see
also \cite{Vir}, \cite{Vir1}). To estimate the value $c_*$, we find the upper
estimate of the number of finite clusters on the
hexagonal lattice  which contain the fixed lattice vertex.

{\bf 2. The percolation theory problem on the hexagonal lattice.}
First of all, we introduce some geometrical objects
modeling crystal lattices. After that we will set the problem
of the discrete percolation theory on such mathematical structures.
In connection with the purpose of the present work, we  consider only
two-dimensional lattices.

We name  the infinite set $V$ in ${\Bbb R}^2$ the periodic one
if there is the pair $\langle {\bf e}_1, {\bf e}_2\rangle$ of
not collinear vectors in ${\Bbb R}^2$ (the parallelogram of
periods) such that the relation $V = V + n_1 {\bf e} _1 + n_2 {\bf
e} _2$ takes place for any $n_i \in {\Bbb Z} $, $i \in \{1,2 \} $.
We name the periodic set in ${\Bbb R}^2$ {\it the crystal
lattice} if it consists of isolated points. The crystal lattice
admits the disjunctive decomposition $V = \displaystyle \bigcup _
{\langle n_1, n_2\rangle \in {\Bbb Z}^2} \{V_0 + n_1 {\bf e} _1 +
n_2 {\bf e}_2 \}$ where the finite set $V_0$ is called {\it the
crystal cell}. If the number of points in $V_0$ is minimal among
all crystal cells admissible in $V$, then such a cell is called
the {\it elementary} one.

Since only the topological structure of the set is important
if the per\-co\-la\-tion of random field is studied then, for the
formulation of per\-co\-la\-tion theory problem, it is   convenient
to use the concept of {\it the periodic graph} and its immersion ${\sf
M}$ into ${\Bbb R}^2$ defining
the connectedness on the crystal lattice.
\smallskip

D e f i n i t i o n\ 1 \cite{1}.\ {\sl Let $ \Lambda = \langle V,
\Phi\rangle $ be the infinite nondirectional loop-free graph where
V is the set of vertexes and $ \Phi $ is the set of some
two-element subsets $\{{\bf x}, {\bf y}\} \subset V$ (edges of the graph).
This graph is called the periodic one
with the dimensionality 2 if it admits such an immersion ${\sf M}
$ into $ {\Bbb R}^2$ when the image ${\sf M} V$ is the crystal
lattice in ${\Bbb R}^2$ and the image ${\sf M} \Phi$ of the set
$\Phi$ is invariant relative to translations with the
parallelogram of periods $\langle {\bf e}_1, {\bf e} _2\rangle$,
$$ {\sf M} \Phi + n_1 {\bf e} _1 + n_2
{\bf e} _2 = {\sf M} \Phi \, \eqno (1)$$ $\langle n_1, n_2\rangle
\in {\Bbb Z}^2$ such that the set $\Phi_0 =\{\{{\bf x}, {\bf y}\}
\in \Phi:{\sf M}{\bf y}\in V_0, {\bf x} \in V\}$ is finite.}
\smallskip

If  $\{{\bf x}, {\bf y}\} \in \Phi$, ${\bf x}$, ${\bf y} \in V$, then
such vertexes are called the adjacent ones and we designate the
adjacency relation between them by means of ${\bf x} \phi {\bf y}$.

Further, we does not distinguish vertexes of the graph and their
images obtained by the immersion ${\sf M}$. We does not
distinguish also immersions in ${\Bbb R}^2$ of the  graph $\Lambda $
differing from each other. Thus, we consider that the vertex set
$V$ of the graph coincides with $\displaystyle \bigcup\limits _
{{\bf x} \in V_0} \{{\Bbb Z} ^2 + {\bf x} \}$ and the property (1)
of its periodicity is written down in the form $ \Phi = \Phi +
n_1 {\bf e}_1 + n_2 {\bf e}_2$. The {\it adjacency} relation
of the periodic graph $\langle V, \Phi\rangle$ is
completely defined by the set $\Phi_0$ since the $\Phi$ admits the
disjunctive decomposition $ \Phi = \displaystyle\bigcup\limits _
{\langle n_1, n_2 \rangle \in {\Bbb Z} ^2} \{\Phi_0 + n_1 {\bf e}
_1 + n_2 {\bf e} _2 \} $.

In this connection, we name the set
$\Phi_0$ the adjacency one. In terms of crystal physics, it
defines some "nearest neighbors" on the crystal lattice of
vertexes being contained in the fixed elementary crystal cell $V_0$.

The infinite periodic two-dimensional graph $\Lambda$ is called
the {\it hexagonal lattice} if its elementary cell $V_0$ contains
two vertexes. Besides, at the choice of the period parallelogram
$\langle 2 {\bf e} _1, {\bf e} _1 + 3 {\bf e} _2/2
\rangle$ defined by basis vectors ${\bf e}_1, {\bf e}_2$ in
${\Bbb R}^2$, it is possible to put $V_0 = \{{\bf x} _1 = {\bf e}
_1 + {\bf e} _2/2, {\bf x} _2 = 2 {\bf e} _1 + {\bf e} _2 \} $ and
$ \Phi_0 = \{\langle {\bf x} _1, {\bf x}_1 - {\bf e} _2\rangle;
\langle {\bf x}_1, {\bf x} _1 + {\bf e}_2/2 - {\bf e} _1\rangle;
\langle {\bf x}_1, {\bf x}_2\rangle; \langle {\bf x} _2, {\bf x}
_2 + {\bf e}_2\rangle; \langle {\bf x} _2, {\bf x} _2 + {\bf e}
_1 - {\bf e}_2/2\rangle \} $. The hexagonal lattice $\langle V,
\Phi\rangle$ is shown on the left-hand side of Fig.1 where the periodic
structure have been represented by the dotted line on the right-hand
side of it. It is formed by shifts of basis vectors of
the plane using the decomposition
on elementary cells which are parallelograms imposed on the lattice.

\begin{figure}[h]
\begin{center}
\centering
\includegraphics[width=15cm, height=6cm]{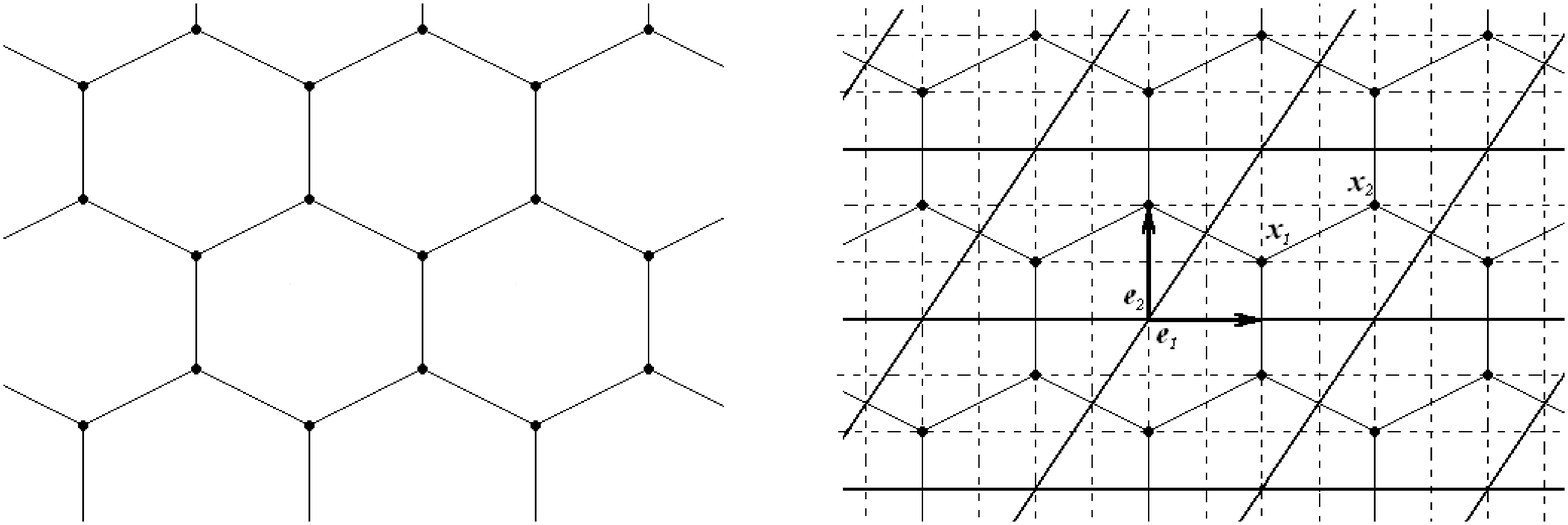}
\caption
{\small The hexagonal lattice.}
\end{center}
\end{figure}

Let us introduced into consideration the Bernoulli random field $
\{\widetilde {c} ({\bf x}); {\bf x} \in V \} $ with the
concentration $c = {\rm Pr} \{\widetilde {c} ({\bf x}) = 1 \}$ on
the graph $ \langle V, \Phi \rangle$. Hereinafter, the tilde which
is put over any mathematical object designates  that it is random.
Each random realization ${\tilde c} ({\bf x}) $, $ {\bf x} \in V $
of the field defines the set $ {\tilde W} = \{{\bf x}: {\tilde c}
({\bf x}) = 1 \} $ that we name the {\it configuration}.
Then, the total set $\{{\tilde c} ({\bf x}); {\bf x} \in V
\}$ of random realizations together with the probability distribution
on them defines the random set on $V$. Its probability
distribution is induced by the probability distribution of the
field $\{{\tilde c} ({\bf x}); {\bf x} \in V \}$. Namely, for
each finite subset $M \subset V$ of lattice vertexes, the
probability of their  filling in the random configuration
${\tilde W}$ is defined by the formula ${\rm Pr} \{M \subset
{\tilde W} \} = c^{| M |}$.

The adjacency relation $\phi$ induces the connectedness for each
random configuration ${\tilde W}$ using the concept of the way on
the graph $\langle V, \Phi \rangle$. The sequence of vertexes
$\langle {\tilde {\bf x}} _0, {\tilde {\bf x}} _1..., {\tilde {\bf
x}} _n\rangle $ chosen from the configuration $ {\tilde W} $ is
called the connected way $\alpha$ with the length $n$ if ${\tilde
{\bf x}}_i \phi {\tilde {\bf x}}_{i +1}$, $i = 0, 1..., n-1$. The
way is called the simple one if $ {\tilde {\bf x}}_i \ne {\tilde
{\bf x}_j}$ in the sequence pointed out at all values $i \ne j$,
$i, j = 1, ..., n$ of indexes and, accordingly, it is called the
cycle if the coincidence of vertexes in the sequence $ \langle
{\tilde {\bf x}}_0, {\tilde {\bf x}} _1..., {\tilde {\bf x}} _n
\rangle $ takes place only at $i = 0$, $j = n$. We name the vertex
pair $ \{{\bf x}, {\bf y}\}$ the {\it connected} one on $ {\tilde
W}$ if $\{{\bf x}, {\bf y} \} \subset {\tilde W}$ and there is a
simple way $\alpha = \langle {\bf x}, {\tilde {\bf x}} _1...,
{\tilde {\bf x}}_ {n-1}, {\bf y} \rangle$ on this configuration.
The connectedness of all vertex pairs is the equivalence relation.
Therefore, any random configuration ${\tilde W}$ is broken up the
family ${\frak M} [{\tilde W}] = \{{\tilde W}_j; j \in {\Bbb N} \}
$ of not intersected and connected sets, ${\tilde W} =
\displaystyle \bigcup_{j = 1}^\infty {\tilde W}_j$, which are
called {\it clusters}. Each cluster consists of vertexes connected
among themselves and any two vertexes being taken from different
clusters are not connected. We designate the cluster of the family
${\frak M}[{\tilde W}]$ which contains the vertex ${\bf x} \in V$
by means of ${\tilde W} ({\bf x})$. If the vertex ${\bf x}$ is not
contained in the configuration ${\tilde W}$ we consider that
${\tilde W} ({\bf x}) = \varnothing$.

The following random function ${\tilde a} ({\bf x})$ describes
the percolation property  of the random field $\{{\tilde c}({\bf
z})$, $ {\bf z} \in V \} $,
$${\tilde a} ({\bf x}) = \left \{\begin {array} {ll} 1;\ \mbox{if}\ |
{\tilde W} ({\bf x}) | = \infty, \\ 0;\ \mbox{if}\ |{\tilde W} ({\bf x}) |
<\infty \, \\
\end {array} \right. $$
where the designation $|\cdot| \equiv {\rm Card} (\cdot) $ is
hereinafter used. In the first case, there is an infinite simple way
$\alpha ({\bf x}) = \langle {\bf x}_i; i \in {\Bbb N}_ + \rangle $
where $ {\bf x} = {\bf x} _0$ and $ {\bf x} _i \in {\tilde W}
({\bf x}) $, $i \in {\Bbb N} _ + $. In the second case, such a way
is absent. On the basis of the random  function ${\tilde a} ({\bf
x}) $, we may define the probability of the percolation from the fixed vertex
${\bf y} \in V$ for the field $\{{\tilde c} ({\bf z}); {\bf z} \in V \}$.
They say that the percolation from this
vertex takes place if the probability $Q (c) = {\rm Pr} \{{\tilde
a} ({\bf y}) = 1 \}$ is positive. This probability does not depend
on the vertex ${\bf y}$ if the periodic graph is uniform (for each pair
${\bf x}, {\bf y} \in V$, there are
such two immersions ${\sf M}_1$, ${\sf M}_2$ in $ {\Bbb R} ^2$ that their images
coincide with each other and ${\sf M}_1 {\bf x} = {\sf M}_2 {\bf y}$).
In the present work, we are
interesting of the value $c_* = \inf \{c : Q (c)> 0 \} $ which is
called the {\it percolation threshold}.

{\bf 3. Finite clusters on the hexagonal lattice.} Following
\cite{1}, we introduce the  concept of the {\it external border}
of the finite cluster ${\tilde W}({\bf x})$. For this aim, we
build another periodic graph $\Lambda^* = \langle V,
\Phi^*\rangle$ on the set $V$ which is called the conjugate one to
the graph $\Lambda$. The adjacency relation $\Phi^*$ on the graph
$\Lambda^*$ is introduced as it is shown on Fig.2 where all
vertexes being $\phi^*$-adjacent with the fixed vertex ${\bf 0}$
are pointed out. They are numbered clockwise. This mean that
vertexes
$$\{{\bf x}_2 + {\bf e}_2, {\bf x}_2 + {\bf e}_1 + 3{\bf e}_2/2,
{\bf x}_2 + {\bf e}_2 + 2{\bf e}_1, {\bf x}_2 + 2{\bf e}_1, {\bf
x}_2 + {\bf e}_1  - {\bf e}_2/2, {\bf x}_2 + {\bf e}_1 - 3{\bf
e}_2/2, \phantom{AA}$$
$$\phantom{A} {\bf x}_2 - 2 {\bf e}_2, {\bf x}_2 - {\bf e}_1 -
3 {\bf e}_2/2, {\bf x}_2 - {\bf e}_1 - {\bf e}_2/2, {\bf x}_2 -
2{\bf e}_1, {\bf x}_2 - 2{\bf e}_1 + {\bf e}_2, {\bf x}_2 - {\bf
e}_1 + 3{\bf e}_2/2\}$$ are $\phi^*$-adjacent with the vertex
${\bf x}_2$ on Fig.1 at the used immersion of
the hexagonal lattice.

\begin{figure}[h]
\begin{center}
\centering
\includegraphics[width=5cm, height=5cm]{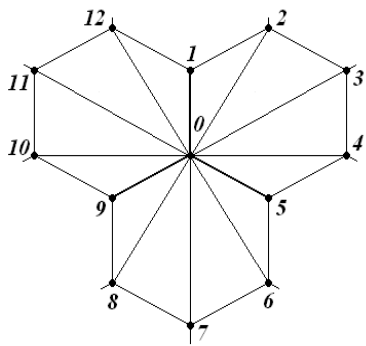}
\caption%
{\small The adjacency relation of the vertex \emph{\textbf{0}}.}
\end{center}
\end{figure}

The adjacency relation $\phi^* $ as well as the relation $\phi$ generates the connectedness for each random configuration ${\tilde W}$. It breaks up both the
configuration ${\tilde W}$ and the configuration $V \setminus
{\tilde W}$ being additional to it into some connected sets.

D e f i n i t i o n\ 2\ \cite {1}. {\sl Let $W$ be the finite
cluster. The set $\partial W$ is called the external border of
$W$ on the configuration ${\tilde W}$ if $W \subset
{\tilde W}$ and $\partial W$
consists of all vertexes ${\bf z} \in {\Bbb Z}^2 \setminus {\tilde W}$
having the following properties.

1. For each ${\bf z}$, there is the vertex ${\bf y} \in W$ such
that ${\bf z} \phi {\bf y}$.

2. For the pointed out vertex ${\bf z}$, there exists the infinite
$\phi$-way $\alpha$ on the graph $\langle V, \Phi \rangle$,
$\alpha \cap W = \varnothing$ which is begun from this vertex and
it is unique in the intersection $\alpha \cap \{{\bf x}\not\in
W:{\bf x}\phi{\bf y},{\bf y}\in W\}$.}

The following statement is valid.
\smallskip

T h e o r e m\ 1. {\sl Let $W({\bf x})$ be the finite cluster
containing the vertex ${\bf x} \in V$. Then $W ({\bf x})$ has the
nonempty finite external border $\partial W ({\bf x})$ which
possesses the following properties.}

1.\ {\sl The set $\partial W ({\bf x}) $ is the simple cycle
on the graph $\langle V, \Phi ^* \rangle $.}

2.\ {\sl The cycle $ \partial W ({\bf x}) $ surrounds the vertex
${\bf x}$ at the periodic immersion $\langle V, \Phi^*\rangle $ into
${\Bbb R}^2$.}
\smallskip

Although this statement is obvious, nevertheless, its
full proof uses the so-called Jordan topological theorem \cite {1}.
\smallskip

The  way simplicity property is inconvenient to express analytically. Therefore,
the estimate proposed below in the item 6 is based on the revealing of suitable
sufficient conditions which are guaranteed that the fixed way forms
the cluster external border. Using these conditions, we will find
the number estimate of external borders of all possible finite clusters
containing the given vertex.

{\bf 4.\ The cluster decomposition on ${\Bbb Z}^2$.}
Let ${\bf A}$ be the
family of finite clusters $W$ containing  the vertex {\bf 0} of the
hexagonal lattice. For each such a nonempty cluster $W$, we define the random event
$A (W) =
\{\tilde W:\, {\bf 0}\in \tilde W,\, W \in {\frak M}[\tilde W]\,,\tilde
W ({\bf 0})= W\}$. The probability of this event is equal to
$${\rm Pr}\{A(W)\} = c^{|W|}(1-c)^{|\partial W|}\,. \eqno(2)$$

According to Theorem 1, for each cluster $W$ of the family ${\bf A}$,  there is
the simple $\phi^*$-cycle beloning to the family
${\bf B} = \{\gamma = \partial W; W \in {\bf A}\}$.
For each $\phi^*$-cycle $\gamma \in {\bf
B}$, we define the event $B(\gamma) = \{\tilde M\,:\, {\bf 0}\in \tilde
M, \tilde W({\bf 0})\in {\frak M}[\tilde W], \partial\tilde W({\bf 0}) =
\gamma\}$ which is represented in the form of the following finite
union of disjoint events
$$B(\gamma) = \bigcup^{}_{W\in {\bf A}\,:\,\partial W = \gamma} A(W)\,.
\eqno(3)$$
The probability $P(\gamma) = {\rm Pr}\{B(\gamma)\}$ is equal to
$$P(\gamma) = \sum^{}_{W\in {\bf A}\,:\,\partial W =\gamma} {{\rm
Pr}\{A (W)\}} = \sum^{}_{W\in {\bf A}\,:\,\partial W =\gamma}
c^{|W|}(1-c)^{|\partial W|}$$
according to (2) and (3).

It is valid the disjoint decomposition
$\{\tilde a({\bf 0}) = 0\} = \displaystyle
\bigcup^{}_{W\in {\bf A}} A(W)$. The family ${\bf A}$ is
decomposed on some nonintersecting classes consisting of clusters
joined by the following property. We refer clusters $W\in {\bf
A}$ to the same class if they have the coinciding external border.
Therefore, it is valid the transformation
$$\bigcup^{}_{W\in {\bf
A}}{...} = \bigcup^{}_{\gamma\in {\bf B}}\
\left(\,\bigcup^{}_{W\in {\bf A}\,:\,\partial W
=\gamma}{...}\right)\,.$$
Further, on the basis of (3), it follows that
$ \{\tilde a({\bf 0}) = 0\} = \displaystyle \left(\bigcup^{}_{\gamma \in {\bf
B}}B(\gamma)\right) \cup \{{\bf 0} \not\in {\tilde W}\} $. Thus, noticing that
Pr$\{{\bf 0} \not\in {\tilde W}\} = 1 - c$ and ${\rm Pr}\{\tilde a({\bf 0})=0\} = 1 - Q(c)$, we come to the following statement.
\smallskip

T h e o r e m\ 2.\ {\sl The probability $1 - Q(c)$ is represented
by the cluster decomposition}
$$1 - Q(c)\ = \ \sum^{}_{\gamma\in {\bf B}}P(\gamma) + 1 - c\,. \eqno(4)$$

The cluster decomposition (4) is converged according to its
definition. But the function $Q(c)$ is differed from zero only at
$c > c_* > 0$ and, therefore, it is not analytic. Some principle
difficulties of the value $c_*$ calculation are connected with
this surcumstance.  At present time, there are not some algorithms
of its evaluation not using the stochastic modeling.

{\bf 5.\ The estimate of percolation probability.} For reception
of the upper estimate of the percolation threshold, it is
necessary to obtain the suitable estimate of the probability
$Q(c)$ from below. Such an estimate is based on the following
statement
\smallskip

L e m m a\ 1. {\sl It is valid the inequality
$$c - Q(c) \le \sum_{n = 3}^\infty (1 - c)^n r_n \eqno (5)$$
where $r_n = |\{\gamma\in {\bf B}\,:\,|\gamma|=n\}|$, $n\geq 3$.}

$ \square $\ We use the elementary upper estimate $P(\gamma) \le (1 - c)^{|\gamma|}$
which follows from Definition 2
and the expression of $B(\gamma)$. Using this estimate and (4),
we come to the upper restriction of the sum
$$\displaystyle
\sum^{}_{\gamma \in {\bf B}}P(\gamma) \leq \sum^{}_{\gamma \in
{\bf B}}{(1 - c)^{|\gamma|}} = \sum^{\infty}_{k=3} {(1 -
c)^{k}r_k}\,.\quad \blacksquare$$
\smallskip

Now, we find the upper estimate of the value $r_n $, $n \ge 3$.
With this aim, we introduce the class ${\bf B}_n$ of all such
simple $\phi^*$-cycles with the length $n$ on the lattice
$\Lambda^*$ that each of them surrounds the point {\bf 0}.
Further, we introduce the set ${\bf C}_n ({\bf x} _0)$ of simple
ways $\gamma = \langle {\bf x} _0, {\bf x} _1..., {\bf x} _n
\rangle $ on the lattice $\Lambda^*$ having the length $n$ and
such that each of them possesses the property ${\bf x}_j \ne {\bf
x}_ {j+2}$, $j = 0, 1..., n-2$. We consider also those subsets
${\bf C}_{n-1} ({\bf x}_0, {\bf x}_1)$ of the last introduced set
${\bf C}_n ({\bf }_0)$ which consist of ways having two leading
fixed vertexes ${\bf x}_0$, ${\bf x}_1$. Besides, each way of
these subsets is such that each two its edges following one after
another, are necessarily a part of the cycle belonging to the
family ${\bf B} = \displaystyle \bigcup\limits_{n=3}^\infty {\bf
B}_n$. In view of the uniformity of the hexagonal lattice (i.e.
the equivalence of all its vertexes), the value $|{\bf C}_{n-1}
({\bf x} _0, {\bf x} _1)|$ does not depend on the vertex ${\bf
x}_0$.

Let $\alpha_k = \langle {\bf x}_0, {\bf x}_1, ..., {\bf
x}_k\rangle$ is the fixed simple way such that it is the part of
circle belonging to the class ${\bf B}$. Further, ${\bf
D}(\alpha_k)$ is the set of edges $\langle {\bf x}_k, {\bf
x}_{k+1} \rangle$ such that ways $\alpha _ {k+1} = \langle {\bf x}
_0, {\bf x} _1..., {\bf x} _k, {\bf x} _ {k+1} \rangle$ possess
the same property as the way $\alpha_k$ has. Denote $n_* =
\max\limits_{\alpha_k}\{|{\bf D}(\alpha_k)|\}$.
\smallskip

L e m m a\ 2. {\sl It is valid the inequality}
$$r_n < n_*(n-2) \max_{{\bf x}_1:\, {\bf x}_1\phi^* {\bf x}_0}
\{|{\bf C}_{n-1}({\bf x}_0, {\bf x}_1)|\}\,. \eqno (6)$$

$\square$\ Let us consider an infinite simple way $\alpha({\bf
0})$ from the vertex ${\bf 0}$ on the lattice $\Lambda$. For
definiteness, this way is chosen in such a manner that each its
finite part has the least length among all ways connecting its
ends. According to Theorem 1, each cycle $\gamma$ on $\Lambda^*$
has necessarily the vertex which is common with $\alpha ({\bf
0})$. Among all such vertexes, we choose the vertex being the
nearest one to the vertex {\bf 0} where the distance is counted
along the way $\alpha ({\bf 0})$.  We designate this vertex by
${\bf z}_\gamma$. Then, each cycle $\gamma \in {\bf B}$ is
contained in the unique class among not intersected classes ${\bf
B} (l)$, $l \in {\Bbb N}$. Each class ${\bf B}(l)$ of the last
family contains only those cycles $\gamma$ of ${\bf B}$ which have
the correspondent vertex ${\bf z}_\gamma$ at the distance $l$
along the way $\alpha ({\bf 0})$, dist\,$({\bf z}_\gamma, {\bf 0})
= l$. On the basis of classes ${\bf B}(l)$, $l \in {\Bbb N}$, we
construct some classes ${\bf B}_n (l) = {\bf B}(l)\cap {\bf B}_n =
\{\gamma \in {\bf B}: | \gamma | = n, {\rm dist} ({\bf z}_\gamma,
{\bf 0}) = l\}$, $n = 3, 4, ...$, $l \in {\Bbb N}$. It is obvious
that the inequality $l \le n-2$ takes place and, therefore, it is
valid ${\bf B} _n = \displaystyle\bigcup\limits_{l=1}^{n-2}{\bf
B}_n (l)$,
$$r_n = \sum\limits_{l=1}^{n-2}|{\bf B}_n (l)|\,. \eqno (7)$$

Let us fix the clockwise direction on cycles $\gamma = \langle
{\bf z}_\gamma, {\bf x}_1...., {\bf x}_ {n-1}, {\bf z}_\gamma
\rangle $ of ${\bf B}_n(l)$. We introduce into consideration
disjoined sets ${\bf B}_n (l, {\bf a}) = \{\gamma \in {\bf B}_n
(l): {\bf x}_1 - {\bf z}_\gamma = {\bf a} \} $ of cycles
containing in ${\bf B} _n (l) $ such that each of them has the
fixed leading vertex that follows after ${\bf z}_ \gamma$
according to the direction on $\gamma$. This vertex is $ {\bf x}
_1 = {\bf z} _ \gamma + {\bf a} $. The vector ${\bf a}$ is one of
vectors on Fig.2 provided that ${\bf z}_\gamma$ is translated to
${\bf 0}$. Then, the disjoint decomposition ${\bf B}_n (l) =
\displaystyle\bigcup\limits_{{\bf y}: {\bf z}_\gamma \phi^* {\bf
y}}{\bf B}_n (l, {\bf y}-{\bf z}_\gamma)$ takes place and,
therefore,
$$|{\bf B}_n (l)| = \sum\limits_{{\bf y}: {\bf z}_\gamma
\phi^* {\bf y}}|{\bf B}_n (l, {\bf y} - {\bf z}_\gamma)|\,. \eqno
(8)$$ It is obvious that $|{\bf B}_n (l, {\bf a})| \le |{\bf C} _
{n-1} ({\bf z}_ \gamma, {\bf z}_\gamma + {\bf a})|$. Namely, any
fixed circle of ${\bf B}_n (l, {\bf a})$ turns to the way of ${\bf
C}_ {n-1} ({\bf z}_ \gamma, {\bf z}_ \gamma + {\bf a})$ after the
removal of the last circle edge according to the order. Thus, we
obtain the injection since no more than one cycle of $ {\bf B} _n
(l, {\bf a})$ corresponds to each way of ${\bf C}_ {n-1} ({\bf
z}_\gamma, {\bf z}_\gamma + {\bf a})$. Thus, on the basis of the
above estimate and (8), it follows the inequality
$$|{\bf B}_n (l)| \le n_*\max_{{\bf x}_1: {\bf x}_1\phi^*
{\bf x}_0}\{|{\bf C}_{n-1}({\bf x}_0, {\bf x}_1)|\}\,.$$ Since the
value $|{\bf C}_{n-1}({\bf x}_0, {\bf x}_1)|$ does not depend on
$l = {\rm dist} ({\bf 0}, {\bf z}_\gamma)$, we obtain (6) applying
the received inequality to do the upper estimate of the right-hand
side  of the decomposition (7). $\blacksquare$
\smallskip

We introduce the construction\ \ for the estimating of the value
$|{\bf C}_{n-1} ({\bf x}_0, {\bf x}_1)|$. We characterize uniquely
each way of ${\bf C}_n ({\bf x}_0)$, $n \ge 2$ by the sequence
$\langle {\bf a}_1, ..., {\bf a}_{n-1} \rangle$ of {\it shift}
vectors where ${\bf a}_1 = {\bf x}_1 - {\bf x}_0$ and each its
vector ${\bf a} _i $, $i = 2..., n $ represents the vector ${\bf
x}_i - {\bf x}_{i-1}$ turned on the angle being least among two
ones formed by the vector ${\bf x}_{i-1} - {\bf x}_{i-2}$ and the
basis vector ${\bf e}_2$ on the lattice plane. Now, we introduce
the set ${\bf G}_n({\bf a}_1)$ of all sequences $\langle {\bf
a}_1, ..., {\bf a}_{n}\rangle$ with the fixed vector ${\bf a}_1$
and  such that each pair $\langle {\bf a} _i, {\bf a} _ {i+1}
\rangle $, $i = 1, 2., n-1$ is admissible in it. Here, the pair
$\langle {\bf a}, {\bf a}'\rangle $ is called the admissible one
if there exists the circle $\gamma$ in the family ${\bf B}$ such
that the corresponding sequence contains the pair $\langle {\bf
a}, {\bf a}'\rangle$ of vectors following one after another in it.
The introduced set is equivalent to the set ${\bf C}_ {n} ({\bf x}
_0, {\bf x} _1)$. Hence, $|{\bf C}_{n} ({\bf x}_0, {\bf x}_1)| = g
({\bf a} _1; n) \equiv | {\bf G} _n ({\bf a} _1)|$.

Let us decompose the set ${\bf G}_{n} ({\bf a}_1)$ by the
following way ${\bf G}_n ({\bf a}_1) =
\displaystyle\bigcup\limits_{{\bf a} _n} {\bf G}_n ({\bf a}_1;
{\bf a}_n)$ on not intersecting sets ${\bf G}_n ({\bf a}_1; {\bf
a}_n)$ of sequences $\langle {\bf a}_1, {\bf a}_2..., {\bf
a}_n\rangle$ where the last shift vector ${\bf a}_n$ is fixed side
by side the first one. Then, $g ({\bf a}_1; n) =
\displaystyle\sum\limits_{{\bf a}_n}| {\bf G}_n ({\bf a}_1, {\bf
a}_n)|$. Introducing the numbering for possible shift vectors
which is presented on Fig.2, we may consider that the value
$\langle |{\bf G} _n ({\bf a} _1, {\bf a} _n) | = g_j({\bf a}_1,
n); j = 1 \div 12\rangle$ is 12-dimensional vector for each $n \in
{\Bbb N}$. Here, the component number $j = 1 \div 12$ is defined
by the number of the shift vector ${\bf a}_n$ in the accepted
numbering.  Thus, $g({\bf a}_1; n) = \displaystyle\sum\limits_{j =
1}^{12}g_j({\bf a}_1; n)$.

Now, let us define the matrix ${\cal S}$ with the dimension equal
to the number of nearest neighbors of the vertex on the conjugate
lattice $\Lambda^*$. For the hexagonal lattice, this number is
equal to 12. Matrix elements $S _ {ij} $, $i, j = 1 \div 12$ may
have values 0 or 1 according to  the following rule. We put
$S_{ij} = 1$ if there is such a pair $\langle {\bf a}_k, {\bf a}_
{k+1}\rangle$ of shift vectors at $k = 1, ..., n-1$ in one of a
sequence in the set ${\bf G}_n ({\bf a} _1,{\bf a} _n)$ for
anything value $n = 3, 4... $ and  vectors ${\bf a} _k, {\bf a} _
{k+1}$ have numbers $i$ and $j$ according to the numbering
accepted on Fig.2. On the contrary, $S_{ij} = 0$ if there is not
the circle with the external border which contains the pointed out
joining of edges. Further, we will name the matrix ${\cal S}$ as
{\it the matrix of way connection}. Since $S_{ij} \ge 0$; $i, j =
1 \div 12$, then its maximal modulo eigenvalue $\lambda_0$ is
positive according to the Frobenius theorem (see, \cite{Gant}) for
matrixes with nonnegative elements.
\smallskip

L e m m a\ 3. {\sl Let the matrix ${\cal S}$ has the unique eigenvalue $\lambda_0 > 0$
with maximal modulus. Then, the asymptotic formula
$$ g_i ({\bf a} _1, n) = C _ {ij} \lambda_0 ^ {n-1} (1 + o
(1))\,, \quad n \to \infty \eqno (9) $$ takes place where the
number $j$ corresponds to the shift vector ${\bf a}_1$ and the
nonzero matrix ${\cal C}$ has non-negative matrix elements $C _
{ij}$.}

$ \square$\ According to the definition of the vector $\langle g_i
({\bf a} _1; n); j = 1 \div 12 \rangle$, the recurrent relation
$$g_i ({\bf a}_1; n) = \sum_{k = 1}^{12} g_k ({\bf a}_1; n-1) S_{ki}$$
takes place for any $n =2, 3... $ and for the vector ${\bf a}_1$
with number $j$. Besides, it takes place $g_i ({\bf a} _1; 1) = S
_{ji}$. Then, using the induction on $n \in {\Bbb N}$, we conclude
that
$$g_i ({\bf a}_1; n) = \left({\cal S}^{n-1}\right)_{ji}\,.$$
Since the matrix ${\cal S}$ has the unique eigenvalue $\lambda_0$
with the maximal absolute value, for the expression in the
right-hand side of last equality, the asymptotic formula (9) is
valid at $n \to \infty$. Besides, $S_{ij}\ge 0$ and, hence, the
nonnegativity of matrix elements $C_ {ij} $ follows from the
Frobenius theorem.\ $\blacksquare$
\smallskip

Now, we may find the below estimate of the probability $Q(c)$.
\smallskip

 T h e o r e m\ 3.\ {\sl Let the maximal eigenvalue $\lambda_0$
of  the matrix ${\cal S}$ of way connections be unique. Then, it
is valid the following below estimate of the probability $Q(c)$}
$$Q(c) \ge c - n_*(1 - c)^2
\sum_{l=3}^\infty (n-2)[(1-c)\lambda_0]^{n-2}\,. \eqno (10)$$

$\square$\ On the basis of Lemmas 1 and 2 and the definition of
the functions $g_i ({\bf a} _1; n)$, $i = 1 \div 12$, we have
$$c - Q(c) \le \sum_{n = 3}^\infty (1 - c)^n r_n \le n_*
\sum_{n = 3}^\infty (n-2)(1 - c)^n \max_{{\bf x}_1}\{|{\bf
C}_{n-1}({\bf x}_0, {\bf x}_1)|\} = $$
$$ = n_* \sum_{n = 3}^\infty (n-2)(1 - c)^n g_{n-1}({\bf a}_1) =
n_* \sum_{n = 3}^\infty (n-2)(1 - c)^n \sum_{i = 1}^{12}g_i({\bf
a}_1; n-1) \,. $$ Applying the asymptotic  formula (9), we obtain
$$c - Q (c) \le n_* C (1 - c) ^2 \sum _ {n = 3} ^ \infty (n-2)
[(1 - c) \lambda_0] ^ {n-2} $$ where the positive constant $C >
\max\limits_j \displaystyle\sum \limits_{i = 1}^{12}C_{ij}$ is
chosen by such a way that the inequality $g_i ({\bf a}_1, n)  < C
\lambda_0^{n-1}$, $n \in {\Bbb N}$ takes place. $\blacksquare$

C o r o l l a r y. {\sl For the percolation threshold $c_*$ of
the Bernoulli random field $\{\tilde c ({\bf x}); {\bf x} \in V \}$
on the hexagonal lattice $ \langle V, \Phi\rangle $, the
inequality $c_*\leq 1 - \lambda ^ {-1} _0$ is valid.}

$\square$\ The series in the right-hand side of the inequality
(10) converges at $(1 - c)\lambda_0 <1$. Then, it takes place at
$c> 1 - \lambda_0^{-1}$. The convergence of this series, applying
the reasoning based on the Borel-Cantelli lemma (see, for example,
\cite {2}) leads to the percolation probability  $Q (c)$ being
distinct from zero. It takes place at the above pointed out
restriction on the parameter $c$. Hence, $c_* \leq 1 -
\lambda_0^{-1}$.\ $\blacksquare$
\smallskip

{\bf 6. The upper estimate of the percolation threshold.} It is
follows from considerations of the previous section that the
problem of the upper estimation of the percolation threshold on
the hexagonal lattice is reduced to calculation of the eigenvalue
$\lambda_0$. For its calculation, first of all, it is necessary to
find the matrix ${\cal S}$. For the evaluation of matrix elements
$S_{ij}$, it is necessary either to find the cluster which has the
external border containing the pair of edges $\langle {\bf a},
{\bf a}'\rangle $ where vectors ${\bf a}$, ${\bf a}'$ have numbers
$i$ and $j$ correspondingly or to prove that there is not the
cluster with such a pair. Following statements describe the
general structure of the matrix ${\cal S}$ and point out its zero
elements.
\smallskip

L e m m a\ 4. {\sl The matrix $S_{ij}$ is symmetric and $S_{ii}=
0$.}

$ \square$\ Since the passage of the cycle being the external
border is possible in both directions, the existence or the
absence of pair edges $\langle {\bf a}, {\bf a}' \rangle $
following one after another in it which have numbers $i$ and $j$
correspondingly leads to existence or absence of the cycle which
has the pair of edges $\langle {\bf a}', {\bf a} \rangle$. It
means that  $S_{ij}= S_{ji}$.

The  equality $S_{ii} = 1$ corresponds to the fact that there is
the pair of edges in the external border which are described by
the pair $\langle {\bf a}, {\bf a}\rangle$ of shift vectors  in
the sequence $\langle {\bf a} _1..., {\bf a} _n\rangle$  and the
vector ${\bf a}$ has number $i$. It means, according to the
description of the cycle by means of the specified sequence that
these edges look as $\langle {\bf x}, {\bf y} \rangle $ and $
\langle {\bf y}, {\bf x} \rangle $. But it is impossible.
Therefore, $S _ {ii} = 0$ for all values $i = 1 \div 12$.
$\blacksquare$
\smallskip

L e m m a\ 5. {\sl The Matrix $S_{ij} $ has the property $S_{1,
j}=S_{1, 14-j}$, $S_{5, j}=S_{5, 10-j}$, $S_{9, j}=S_{9, 18-j}$,
$j=2\div 12$ where the subtraction operation in bottom indexes is
understood modulo 12.}

$\square$\ The first relation follows from the symmetry of Fig.2
describing the joining of edges in the vertex {\bf 0}  relative
to the reflection from the straight line defined by vertexes ${\bf
1}, {\bf 0}, {\bf 7}$. The second relation follows from the
symmetry of this figure to the reflection from the straight line
passing through vertexes ${\bf 5}, {\bf 0}, {\bf 11}$ and the
third relation is connected with the reflection from the straight
line passing through vertexes ${\bf 3}, {\bf 0}, {\bf 9}$.
$\blacksquare$
\smallskip

L e m m a\ 6. {\sl Matrix elements $S _ {ij} $ have the property
$S _ {i+4, j+4} = S_{ij}$ where sums $i+4$ and $j+4$ are
understood modulo 12.}

$\square$\ This property follows from the symmetry of Fig.2
describing all possible connections of edges in the vertex {\bf 0}
relative to rotations on the angle $2\pi/3.$\ $\blacksquare $
\smallskip

C o r o l l a r y.\ {\sl The matrix ${\cal S} $ has the following
block structure made of two $4\times 4$-matrices $F, G$}
$$S = \left(\begin{array}{lll} F & G & G^+ \\ G^+ & F & G \\ G & G^+ & F \\
\end{array}\right)\,. \eqno (11)$$

$\square$\ Using $4 \times 4$-matrices $F, G, H $ with matrix elements $F _
{ij} = S _ {ij} $, $G _ {ij} = S _ {i, j+4} $, $H _ {ij} = S _ {i,
j + 8} $, $i, j = 1 \div 4$, it follows from Lemma 6 that
the matrix $ {\cal S} $ is represented as
$${\cal S} = \left( \begin{array}{lll} F & G & H \\ H & F & G \\
G & H & F \\ \end{array}\right)\,.$$
From Lemma 4, it follows that $H = G^+ $, as $G_{ij} = S_{i, j + 4} = S_{i + 8, j}$
at $i, j = 1 \div 4$ and, simultaneously, $S_{i + 8, j} = S_{j, i + 8} = H^+ $.
$\blacksquare$
\smallskip

L e m m a\ 7. {\sl $S_{1,5} = 0$.}

$\square$ Each vertex ${\bf x}$ of the external border $\partial W$
should have the vertex ${\bf y}$ adjacent with it and belonging to
$W$. Besides, there exists the vertex ${\bf z}$ adjacent with ${\bf x}$,
but not belonging $W\cup \partial W$. If we admit that $S _ {1,5} = 1$,
then vertexes {\bf 1} and {\bf 5} (see, Fig.3 where joinings of edges in the vertex
{\bf 0} are shown by the dotted line) belong to $\partial W$ and, hence, they do not
belong to $W$. Since the vertex {\bf 0} has three vertexes {\bf 1},{\bf 5},{\bf 9}
adjacent with it, then
the vertex {\bf 9} should be simultaneously the vertex of $W$ and it does not
belong to $W \cup \partial W$. But, it is impossible. $ \blacksquare$

\begin{figure}[h]
\begin{center}
\centering
\includegraphics[width=7cm, height=7cm]{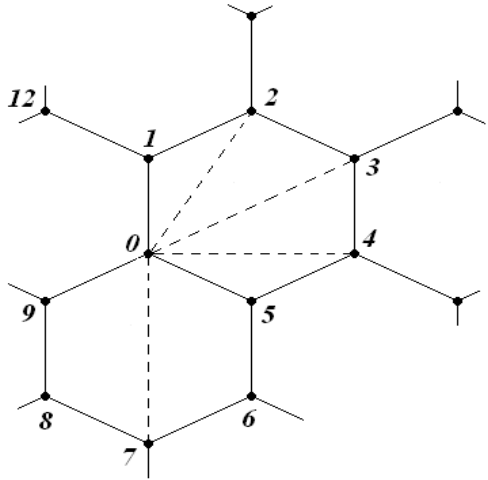}
\caption%
{\small Possible joinings of edges in the vertex \emph{\textbf{0}}.}
\end{center}
\end{figure}

L e m m a\ 8. {\sl $S_{ij} = 0$, $i,j = 1\div 4$.}

$\square$ It is sufficient to prove that $S _ {1, j} = 0$, $j = 2,
3, 4$ and $S _ {2, j} = 0$, $j = 3,4$. We consider the first group
of matrix elements. Vertexes {\bf 0} and {\bf 1} belong to
$\partial W$. There are vertexes ${\bf x}, {\bf y} \in W$ such
that ${\bf x} \phi {\bf 1}$, $ {\bf y} \phi\, {\bf 0}$ and also
there are vertexes ${\bf x}', {\bf y}'\not\in W \cup \partial W$
such that ${\bf x}' \phi \, {\bf 1}$, ${\bf y}' \phi\, {\bf 0}$.
Further, there are ways from vertexes ${\bf x}'$ and ${\bf y}'$ to
infinity not crossing $W$. Vertexes {\bf 2} and {\bf 12} may play
the role of vertexes ${\bf x}, {\bf x}'$ for the vertex {\bf 1}
and only vertexes {\bf 5} and {\bf 9} may be the same for the
vertex {\bf 0}. Then, there are four choice variants of vertex
pairs ${\bf x}, {\bf x}'$ and ${\bf y}, {\bf y}'$ among presented
possibilities. It is sufficient to analyze only two cases which do
not reduce to each other by mirror reflection relative to the
straight line passing through vertexes {\bf 1},{\bf 0},{\bf 7}. We
choose the following possibilities. Vertexes {\bf 2} and {\bf 12}
are the vertexes ${\bf x}, {\bf x}'$ in both cases.  And the
vertexes {\bf 9} and {\bf 5} are the vertexes $ {\bf y}, {\bf y}'$
in the first case and, on the contrary, they are interchanged
their position in the second.

In the first case, there are not infinite ways from vertexes {\bf
1} and {\bf 0} which are crossed with $W$ and passed through
vertexes {\bf 12} and {\bf 5} correspondingly. We consider the
line in the plane where the lattice is placed. This line consists
of the first infinite way of lattice edges that comes from
infinity to the vertex {\bf 12}. Further, it consists of the
sequence of edges $\langle {\bf 12}, {\bf 1} \rangle, \langle {\bf
1}, {\bf 0} \rangle, \langle {\bf 0}, {\bf 5} \rangle$ and it is
ended by the second infinite way from the vertex {\bf 5} to
infinity. The constructed line divides the plane into two parts so
that vertexes ${\bf x}$ and $ {\bf y}$ are in different ones.
Then, vertexes ${\bf x}$ and $ {\bf y} $ should not belong to the
same cluster $W$ since any way from the vertex ${\bf x}$ to the
vertex ${\bf y}$ on lattice edges crosses necessarily this line.
Besides, it may be done only in the lattice vertex. Hence, the
edge connection $\langle {\bf 1}, {\bf 0} \rangle$ with one of
edges $\langle {\bf 0}, {\bf j}\rangle$, ${\bf j} = {\bf 2}, {\bf
3}, {\bf 4}$ is impossible in the variant under consideration.

In the second case, it is considered the analogous line on the
plane which consists of the way coming from infinity to the vertex
{\bf 1} without crossing of the cluster $W$. Further, it consists
of the edge $\langle {\bf 1}, {\bf 0}\rangle$, the diagonal of the
given lattice hexagon which is led from the vertex {\bf 0} to the
vertex ${\bf j}$ (it is presented on Fig.3 where this side has the
border vertexes ${\bf 0},{\bf 1},{\bf 2},{\bf 3},{\bf 4},{\bf 5}$)
and the way from the vertex ${\bf j}$ to infinity on lattice edges
without the crossing of $W$. The line constructed divides the
plane into two parts so that vertexes ${\bf x}$ and ${\bf y}$ are
in different ones. Therefore, they should not belong to the same
cluster $W$ since any way on lattice edges from the vertex ${\bf
y}$ to the vertex ${\bf x}$ necessarily crosses the line in the
lattice vertex.

Contradictions obtained in both cases show that $S_{1, j} = 0$ at $j = 2, 3, 4$.

Let us consider the second group of matrix elements $S_{23}$,
$S_{24}$. We construct the following line on the lattice plane.
Firstly, it consists of the way which comes to the vertex ${\bf
2}$ from infinity without crossing of the cluster $W$. It should
exist since the vertex ${\bf 2}$ belongs to $\partial W$. Further,
the line consists of the consecutive passage of two diagonals on
the hexagon presented on Fig.3. Diagonals are defined by vertex
pairs $\langle {\bf 2}, {\bf 0} \rangle$ and $\langle {\bf 0},
{\bf j}\rangle$ where ${\bf j} = {\bf 3},{\bf 4}$. The line is
ended by the way from the vertex ${\bf j}$ to infinity. It divides
the plane into two parts and the cluster $W$ should be settled
down completely in one of them.

It is obvious that $S_{23} = 0$ since the cluster $W$ may be not
settled down on the right-hand side of the constructed line
according to the direction accepted on it. Otherwise, the vertex
{\bf 0} has no vertexes belonging to $W$ and being adjacent with
it. On the other hand, the cluster $W$ may not be on the left-hand
side of this line since there is the way leaving from the vertex
{\bf 0} to infinity  without crossing of the cluster $W$. It
should exist according to the definition of the external border
vertex. This way should be also in the left part of the plane.
Then, the way divides the left-hand side into two parts again. The
cluster $W$ should be settled down completely in one of them and,
therefore, one of vertexes {\bf 2} or ${\bf j}$ does not belong to
the external border since there is not anything vertex of $W$
adjacent to it. The obtained contradiction proves the equality
$S_{24} =0$.$ \blacksquare$
\smallskip

T h e o r e m\ 4. {\sl The matrix ${\cal S}$ is presented by the formula
$$S = \left(\begin{array}{lll}
{\bf 0}   & G   & G^+ \\
G^+ & {\bf 0}   & G \\
G   & G^+ &  {\bf 0} \\ \end{array}\right)\,,\quad G =
\left(\begin{array}{llll}
0 & 1 & 1 & 1 \\
0 & 1 & 1 & 1 \\
0 & 1 & 1 & 1 \\
0 & 1 & 1 & 1 \\
\end{array}\right) \eqno (12)$$
where ${\bf 0}$ is the zero matrix.}

$ \square $ It follows from Lemmas 5 and 8 that $S _ {3,4} = S _
{3,2}= S_{2,3} =0$. Then the matrix $F$ in the formula (11) is
zero owing to Lemma 4. On the basis of the property $S_{1, j} =
S_{1, 14 - j}$ and Lemmas 7 and 8, we find
$$S_{1,12} = S_{1,2} = 0; \quad S_{1,11} =
S_{1,3} = 0; \quad S_{1,10} = S_{1,4} = 0; \quad S_{1,9} = S_{1,5}
= 0\,.$$
Then, the first line in the matrix $G^+$ in the formula (11) and, hence, the first
column in the matrix $G$ are zero. Other elements of the matrix $G$ are equal to
unity due to the above mentioned criterion at the evaluation of
matrix elements. So, it is had $S_{ij} = 1$ at $i = 1 \div 4, j = 6,
7, 8$ (see Figs. 4-7). $\blacksquare$

\begin{figure}[h]
\begin{center}
\centering
\includegraphics[width=15cm, height=5cm]{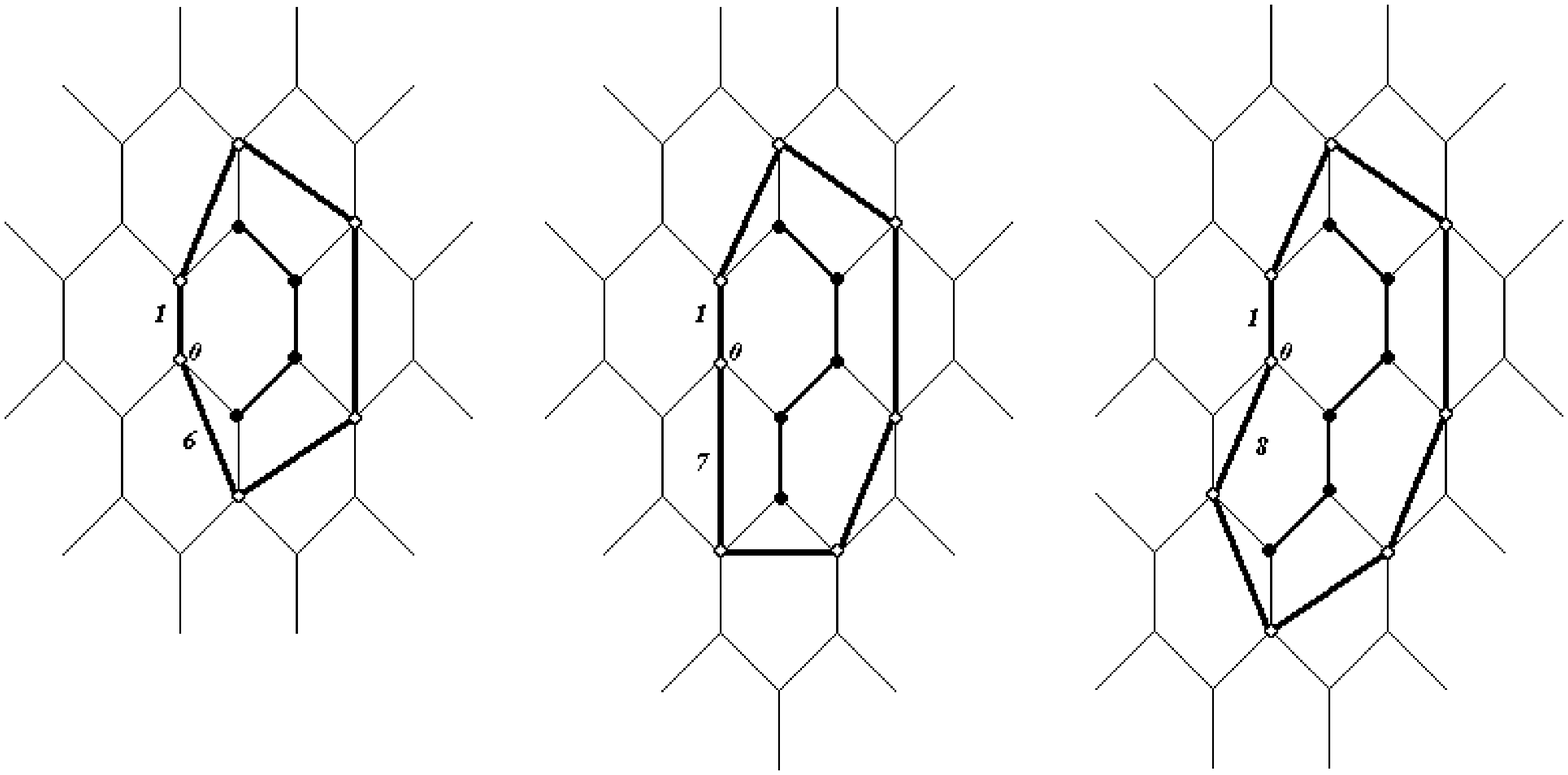}
\caption%
{\small Clusters containing the edge \emph{\textbf{1-0}}.}
\end{center}
\end{figure}

\begin{figure}[h]
\begin{center}
\centering
\includegraphics[width=15cm, height=5cm]{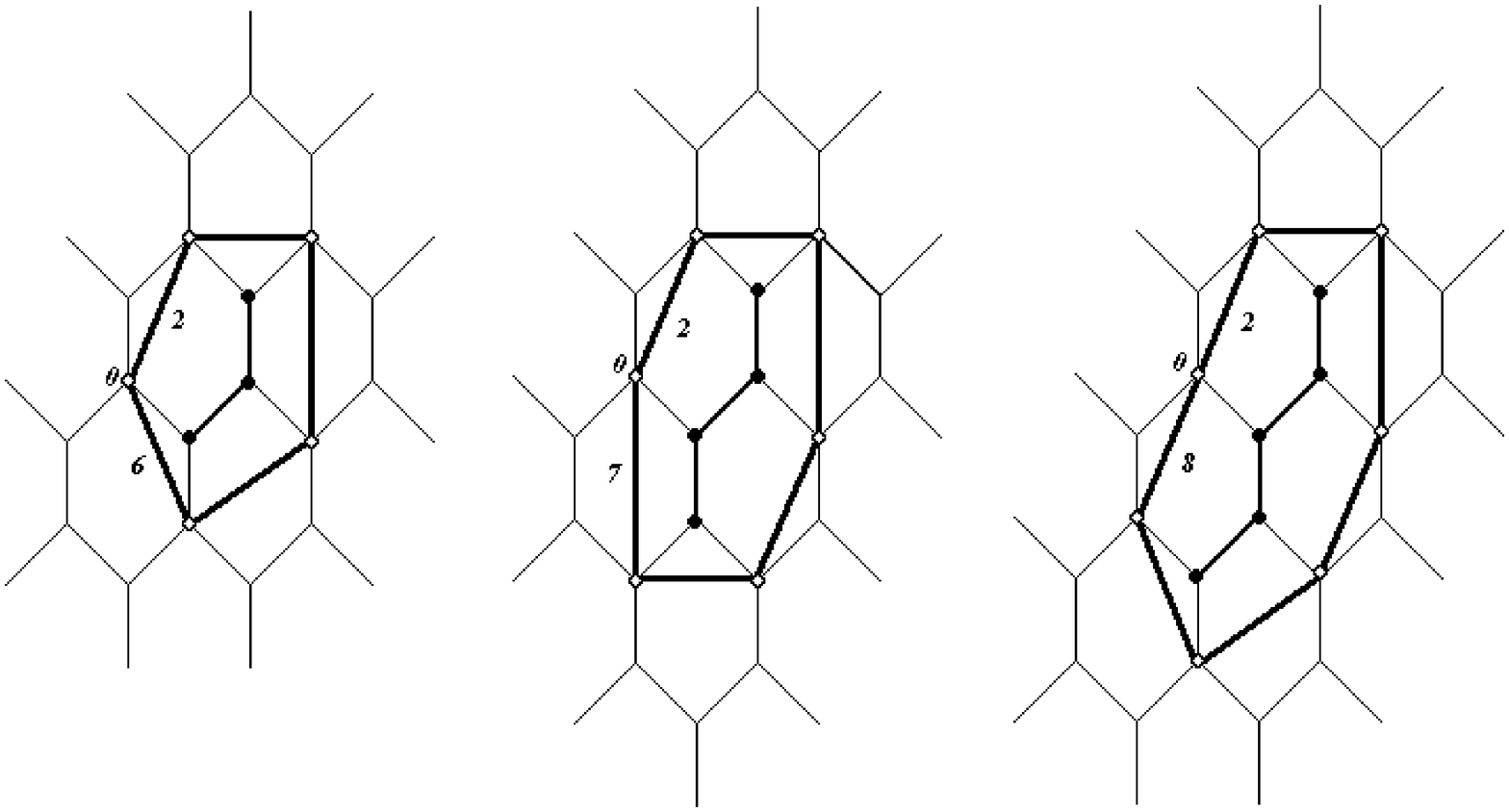}
\caption%
{\small Clusters containing the edge \emph{\textbf{2-0}}.}
\end{center}
\end{figure}

\begin{figure}[h]
\begin{center}
\centering
\includegraphics[width=15cm, height=5cm]{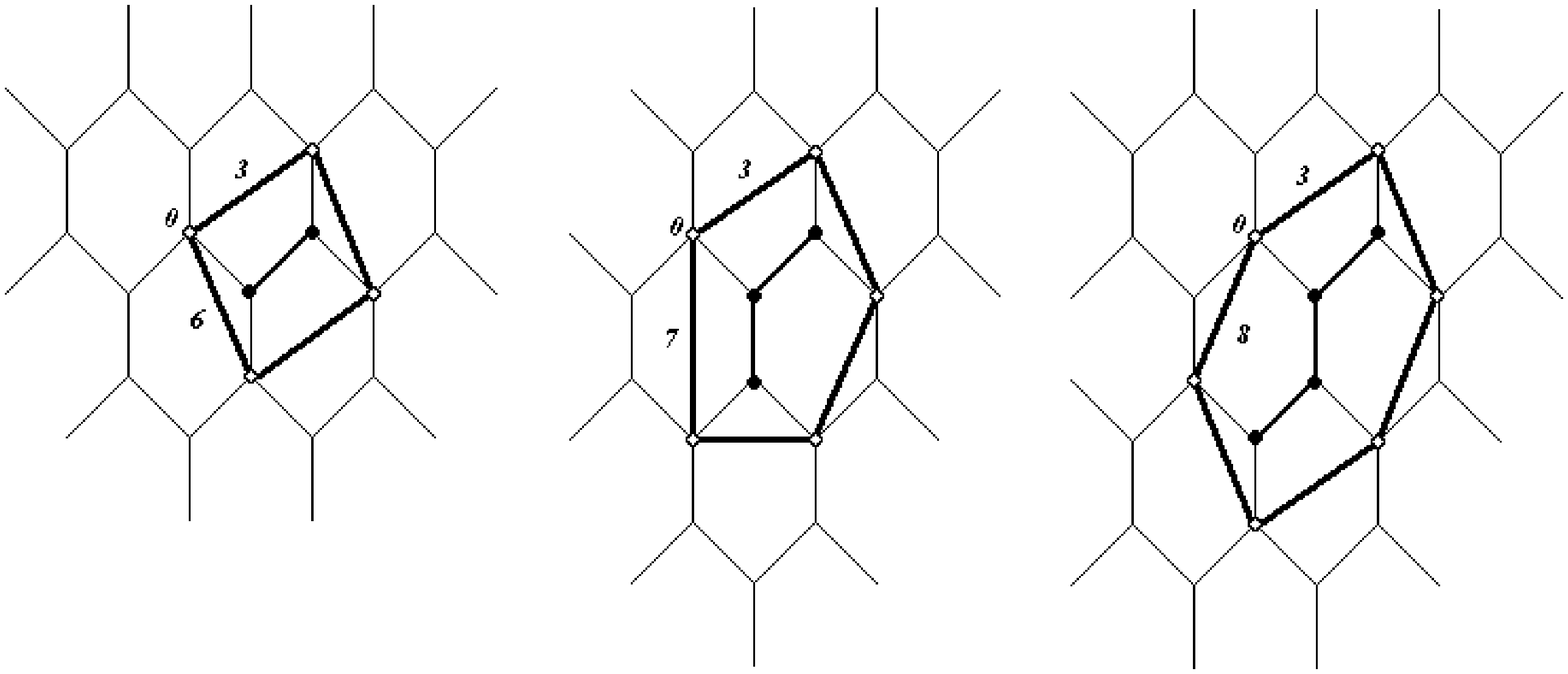}
\caption%
{\small Clusters containing the edge \emph{\textbf{3-0}}.}
\end{center}
\end{figure}

\begin{figure}[h]
\begin{center}
\centering
\includegraphics[width=15cm, height=5cm]{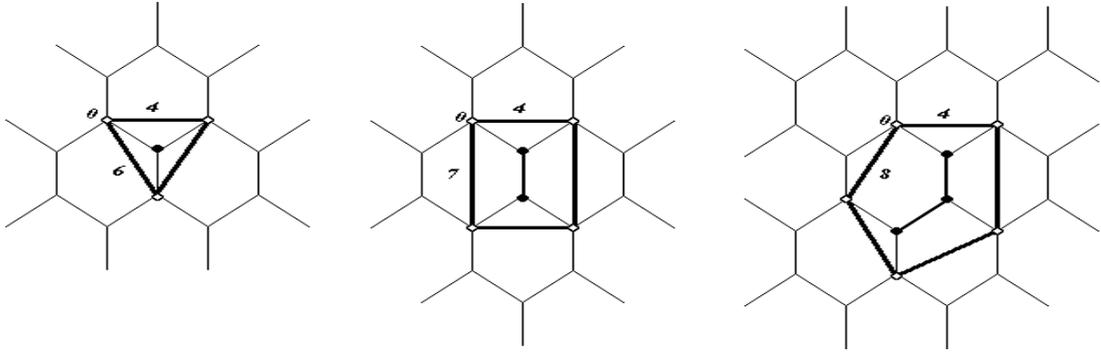}
\caption%
{\small Clusters containing the edge \emph{\textbf{4-0}}.}
\end{center}
\end{figure}

N o t i c e.\ It follows from the matrix ${\cal S}$ expression that $n_* = 7$.
\smallskip

On the basis of Theorem II in Appendix, having put $n = 3$, $m
=4$, $A = {\cal S}$, $A^{(1)} = {\bf 0}$, $A^{(2)} = G$, $A
^{(3)} = G^+$, we conclude that the maximal eigenvalue $\lambda_0$ of the matrix
${\cal S}$ coincides with the maximal eigenvalue of the
matrix
$$B = {\bf 0} + G + G^+ = \left(\begin{array}{llll}
0 & 1 & 1 & 1 \\
1 & 2 & 2 & 2 \\
1 & 2 & 2 & 2 \\
1 & 2 & 2 & 2 \\
\end{array}\right)$$
The rank of this matrix is obviously equal 2. Then, it has
two zero eigenvalues and, therefore, its characteristic equation looks as follows
$$\det(B - \lambda {\bf 1}) = (\lambda^2 - \lambda \xi_1  + \xi_2)
\lambda^2 = 0$$
where $\xi_1 = {\rm Sp}\,B$, $\xi_2 = [{\rm
Sp}\,B^2 - ({\rm Sp}\,B)^2]/2$. Then, $\xi_1 = {\rm Sp}\,B = 6$
and $\xi_2 = 3$ since diagonal elements of the matrix $B^2$ are equal $(B^2)_
{ii} = 3, 13, 13, 13$. Hence, the number
$\lambda_0$ is the greatest root of the quadratic equation $\lambda^2 - 6
\lambda-3 =0$. Whence, we obtain $\lambda_0 = 3 + 2 \sqrt {3}$.
Thus, we have proved the following statement.
\smallskip

T h e o r e m\ 5.\ {\sl The maximal eigenvalue $\lambda_0$ of the matrix
${\cal S}$ is equal $3 + 2 \sqrt {3} \approx 6,46$.}
\smallskip

From Theorem 5 and Corollary of Theorem 3, it follows directly

{\bf The basic statement.}\ {\sl The percolation threshold $c_*$
of the Bernoulli field on the hexagonal lattice does not surpass
the number $2/3(3 -\sqrt{3})$.}

\newpage

\noindent {\bf Appendix}
\medskip

L e m m a\ I.\ {\sl Let $B$ be the $m \times m$-matrix with
nonnegative elements such that there is the number $\mu > 0$ and
the vector ${\sf b} \in {\Bbb R}^m$, ${\sf b} = \langle
b_1,...,b_m\rangle$ with nonnegative components such that it takes
place the inequality}
$$\mu b_k \le \sum_{l=1}^m B_{kl} b_l\,, \quad k = 1 \div m\,.\eqno (
A1)
$$ {\sl Then the maximal eigenvalue $\lambda (B)$ of the matrix $B$ on
the absolute value  is positive and it satisfies the inequality
$\lambda (B) \ge \mu$.}

$\square$\ We apply $N$ times the inequality (A1). Using the induction on $N$, it is
obtained the inequality
$$\mu^N b_k \le \sum_{l=1}^m \left(B^N\right)_{kl}b_l\,. \eqno (A2)$$
At first, we prove the inequality (A1) in the case when the matrix
$B$ has only one-dimensional eigenspaces. Let ${\sf c}^{(r)}=
\langle c_k^{(r)}; k = 1 \div m\rangle$ be eigenvectors of the
matrix $B$ with correspondent eigenvalues $\lambda_r$, $r = 1 \div
m$. We consider that all $\lambda_r$ have been ordered according
to the decrease of their absolute values (at the coincidence of
absolute values, they are ordered according to their phases). We
write down the decomposition of the vector
$${\sf b} = \sum_{r = 1}^m \xi_r {\sf c}^{(r)},\quad \xi_r \in {\Bbb C}\,,
r = 1 \div m\,.$$ Then,
$$B^N {\sf b} = \sum_{r = 1}^m \xi_r \lambda_r^N {\sf c}^{(r)}$$
and, therefore, the inequality (A2) is represented in the form
$$\mu^N b_k \le \sum_{r = 1}^m \xi_r \lambda_r ^N {\sf c}^{(r)}_k\,.
\eqno (A3)$$
There is the number $k \le m$ such that $b_k > 0$.
Let us consider the inequality (A3) for this number. Among all numbers $r = 1
\div m$, there is the minimal number $r_0$ such that
$\xi_{r_0}c^{(r_0)}_k \ne 0$. Then, the inequality (A3) is rewritten in the form
$$\mu^N b_k \le \sum_{r = r_0}^m \xi_r \lambda_r ^N {\sf c}^{(r)}_k$$
or, calculating the $N$-th degree root for both positive parts of the inequality, we
obtain
$$\mu b_k^{1/N} \le \lambda_{r_0}
\left[\sum_{r = r_0}^m \xi_r (\lambda_r/\lambda_{r_0})^N {\sf
c}^{(r)}_k\right]^{1/N} = |\lambda_{r_0}| \left|\sum_{r = r_0}^m
\xi_r (\lambda_r/\lambda_{r_0})^N {\sf
c}^{(r)}_k\right|^{1/N}\,.$$ The last equality is valid due to the
positivity of the right-hand side of (A3). Further, we go to the
limit $N \to \infty$. In this case, $\displaystyle\lim\limits_{N
\to \infty}b_k^{1/N} = 1$. Due to the boundedness of the summing
expression in the right-hand side of the inequality, its upper
limit does not surpass the unity. Then, $\mu \le |\lambda_{r_0}|$.
According to the definition and due to the Frobenius theorem,
$|\lambda_{r_0}| \le  \lambda(B)$.

One may change any matrix $B$ by such a way that all its
eigenspaces turns into one-dimensional ones. It is done by means
of addition to the matrix $B$ of the suitable matrix with
nonnegative elements. It may be done as much near to zero matrix
as one wants. This nearness is understood according to the ${\Bbb
R}^{m^2}$ topology. The possibility of this follows from the fact
that the multiplicity condition of eigenvalues satisfying the
characteristic equation $\det (B - \lambda {\bf 1}) = 0$ is
expressed by the supplement equation $\displaystyle \frac d {d
\lambda} \det(B - \lambda {\bf 1}) = 0$. The last equation cuts
out a differential manifold with the codimensionality one in the
space ${\Bbb R}^{m^2}$ of admissible matrices $B$. At such an
addition of small matrix, the matrix $B$ transforms to such a
matrix $B_\epsilon$ that $B_\epsilon \to B$ at $\epsilon \to +0$.

The inequality (A1) is proved in general case by the following
way. Let ${\sf b}$ be the eigenvector of the matrix $B$
corresponding to $\lambda(B)$. Since
$$\lambda (B) b_k = \sum_{l=1}^m B_{kl}b_l \le \sum_{l=1}^m
\left(B_\epsilon\right)_{kl}b_l$$ and the matrix $B_\epsilon$ has
only one-dimensional eigenspaces, the inequality (A1)  takes place
where it is needed to change both $\lambda(B)$ on $\lambda
(B_\epsilon)$ and $\mu$ on $\lambda (B)$. Consequently, $\lambda
(B) \le \lambda (B_\epsilon)$. After that, we come to the limit
$\epsilon \to + 0$. $\blacksquare$

T h e o r e m\ I. {\sl Let $A$ be the $n \times n$-matrix with
nonnegative elements. It consists of $m^2$ blocks being
rectangular matrixes $B^{(k,l)}$ which have correspondingly $p_k$
lines and $s_l$ columns $k,l = 1 \div m$, $m \le n$, $p_1 + p_2 +
... + p_m = n$, $s_1 + s_2 + ... + s_m = n$,}
$$A = \left(\begin {array}{llll} B^{(1,1)} & B^{(1, 2)} & \dots & B^{(1, m)}\\
B^{(2,1)} & B^{(2, 2)} & \dots & B^{(2, m)}\\
\vdots & \vdots & \vdots & \vdots\\
B^{(m,1)} & B^{(m, 2)} & \dots & B^{(m, m)}\\
\end {array}\right)\,. \eqno (A4)$$
{\sl Matrix elements of the square $m\times m$-matrix $B$ are defined by the formula
$$B_{kl} = \max\left\{\sum_{j = 1}^{s_l} B^{(k, l)}_{ij}; i = 1\div p_k
\right\}\,,  \quad k,l = 1\div m\,.$$ Then, positive eigenvalues
$\lambda (A)$ and $\lambda (B)$ corresponding to matrixes $A$ and
$B$ which are maximal on their absolute values satisfy the
inequality $\lambda (B) \ge \lambda (A)$.}

$\square$ Matrixes $A$ and $B$ possess nonnegative elements. Due
to the Frobenius theorem \cite{Gant}, eigenvalues of these
matrixes are positive if they are maximal on their absolute
values. Besides, due to this theorem, there are such eigenvectors
of matrixes  $A$ and $B$ corresponding to eigenvalues pointed out
that they have nonnegative components. We designate them as
follows ${\sf a} = \langle a_1, ..., a_n\rangle$, $a_j \ge 0$, $j
= 1 \div n$; ${\sf b} = \langle b_1, ..., b_m\rangle$, $b_i \ge
0$, $i = 1 \div m$. Then,
$$\sum_{j=1}^n A_{ij}a_j = \lambda (A)a_i\,, \quad
\sum_{l=1}^m B_{kl}b_{l} = \lambda (B)b_k\,.$$
We introduce  the vector ${\sf c} \in {\Bbb R}^m$,
${\sf c} = \langle c_1, ..., c_m \rangle$ with components
$c_k = \max\{a_i; s_1 + ... + s_{k-1} < i \le s_1 + ... + s_k\}$.
Dividing summations on $j = 1 \div n$ in both sides of
first equality, we introduce repeated summations. First of them is done
on groups containing $s_1, ..., s_m$ numbers, the second is done within each
of these groups. In a result, we obtain
$$\sum_{j = 1}^n A_{ij}a_j =
\sum_{l = 1}^m  \sum_{j= s_1 + ... + s_{l-1}+1}^{s_1 + ... + s_l} A_{ij} a_j \le
\sum_{l = 1}^m c_l \sum_{j= s_1 + ... + s_{l-1}+1}^{s_1 + ... + s_l}A_{ij}\,.$$
Using the change $i \Rightarrow   p_1 + ... + p_{k-1}
+ i$, $i = 1 \div p_k$, we rewrite the last inequality in the form
$$\lambda (A)a_{i + p_1 + \dots + p_{k-1}}
\le \sum_{l = 1}^m c_l\sum_{j = 1}^{s_l} B_{i, j}^{(k,l)}\,.$$
Calculating the maximum on $i$ in both sides of the inequality, we
obtain
$$\lambda(A)c_k \le \sum_{l=1}^m B_{k, l} c_l\,. \eqno (A5)$$
The last inequality coincides with (A1), if we put $\mu = \lambda
(A)$ and change ${\sf b}$ on ${\sf c}$. Then, the theorem
statement follows from (A1). $\blacksquare$
\bigskip

T h e o r e m\ II. {\sl Let $\langle A^{(1)}, ..., A^{(n)}
\rangle$ be the ordered collection of $m\times m$-matrixes with
nonnegative elements. Further, let $nm \times nm $-matrix
$A$ be made up of $m \times m$-matrixes $B^{(k, l)}$ according to
the following formula}
$$A = \left( \begin{array}{llll}
B^{(1, 1)} & B^{(1, 2)} & \dots & B^{(1, n)}\\
B^{(2, 1)} & B^{(2, 2)} & \dots & B^{(2, n)} \\
\vdots & \vdots & \vdots & \vdots \\
B^{(n, 1)} & B^{(n, 2)} & \dots & B^{(n, n)} \\
\end{array}\right)\,, \eqno (A6)$$
{\sl where each of ordered collections $\langle B^{(i, 1)}, ...,
B^{(i,n)} \rangle$, $i = 1, ..., n$ is obtained by the permutation
${\sf P}\in {\Bbb P}_n$ of the collection $\langle A^{(1)}, ...,
A^{(n)}\rangle$ and ${\Bbb P}_n$ is the permutation group of the
$n$-th order. Then, the maximal eigenvalue of the matrix $A$
coincides with the maximal eigenvalue of the $n \times n$-matrix
$B = A^{(1)} + ... + A^{(n)}$.}
\smallskip

$\square$\ The matrix $A$ has nonnegative elements. Therefore,
according to the Frobenius theorem, its eigenvalue $\lambda (A)$ being maximal on the absolute value is positive. Let ${\sf a}\in {\Bbb R}^{nm}$, ${\sf
a} = \langle a_1, ..., a_{nm}\rangle$ be the eigenvector which corresponds to this eigenvalue where $a_i \ge 0$, $i = 1 \div nm$, $A{\sf a} = \lambda (A) {\sf a}$. It means that
$$\sum_{j = 1}^{nm} A_{ij} a_j = \lambda (A) a_i\,,\quad i = 1 \div nm\,.$$
We define the vector ${\sf c} \in {\Bbb  R}^m$, ${\sf c} = \langle
c_1, ..., c_m\rangle$, $c_j \ge 0$, $j = 1 \div m$ where
$$c_k = \max\{a_{k + sm}; s = 0, \dots , n-1\}\,, \quad k = 1 \div m\,.$$
Then, changing $i = k + pm$, $j = l + sm$, $s, p = 0, 1, ...,
n-1$, we have
$$\lambda (A)a_{k + pm} =
\sum_{s = 0}^{n-1} \sum_{l = 1}^{m}A_{k + pm, l + s m} a_{l + sm}
\le \sum_{l = 1}^{m} \left(\sum_{s = 0}^{n-1} A_{k + pm, l + s
m}\right) c_{l}\,.$$
Calculating the maximum on $p = 0, 1, \dots,  n-1$ of both sides of the inequality, we obtain
$$\lambda (A) c_k \le \sum_{l=1}^m B_{kl} c_l$$
where
$$B_{kl} = \max\left\{\sum_{s = 0}^{n-1}A_{k + pm, l + sm}; p = 0, 1, ...,
n-1\right\} \eqno (A7)$$
are matrix elements of $B = A^{(1)} + ... + A^{(n)}$. This formula follows
 from the equality
$$\sum_{s = 0}^{n-1} A_{k +pm, l + sm} =
\sum_{s = 0}^{n-1}\left(A^{({\sf P}(s+1))}\right)_{kl}$$
where the permutation ${\sf P} \in {\Bbb P}_n$ is defined by the number $p$
being the line number in the block matrix (A6). Having changed the summation variable
as follows ${\sf P}(s + 1) \Rightarrow s$, we find that the last expression is equal to
$$\sum_{s = 0}^{n-1}\left(A^{({\sf P}(s +1))}\right)_{kl} =
\sum_{s = 1}^n A^{(s)}_{kl} = B_{kl}$$ and the sum in (A4) does not depend on $p$.

Applying the statement of Theorem I to the matrix $A$, in a
result, we obtain the inequality $\lambda (B) \ge \lambda (A)$. On
the other side, we take the eigenvector    ${\sf b}$ of the matrix
$B$ with nonnegative components that corresponds to the eigenvalue
$\lambda (B)$. Further, we define the vector ${\sf c}' = \langle
\underbrace{{\sf b}, {\sf b}, \dots, {\sf b}}_{n\,} \rangle$. For
this vector, we have
$$A {\sf c}' = \langle B {\sf b}, B {\sf b}, \dots, B{\sf b}\rangle =
\lambda (B) {\sf c}'\,.$$
Then,  $\lambda (B)$ is the eigenvalue of the matrix $A$ with the eigenvector
${\sf c}'$. According to the definition of eigenvalue $\lambda (A)$, we have $\lambda (A) \ge \lambda (B)$. From two obtained inequalities, it follows that $\lambda (A) =
\lambda (B)$.\ $\blacksquare$

\end{document}